\documentclass[conference]{IEEEtran}

\usepackage{amsmath,amssymb,eucal,graphicx}
\usepackage{pst-plot}
\usepackage{epsfig}
\usepackage{amsthm}
\usepackage{amsfonts}
\usepackage{epsfig}
\usepackage{float}
\usepackage{pstricks}
\usepackage{color}
\usepackage{multicol,subfigure}
\usepackage{mathcomSTEP,stmaryrd}
\usepackage{balance}

\theoremstyle{remark}

\theoremstyle{definition}

\begin{document}

\title{
Primary Rate-Splitting Achieves Capacity for the Gaussian Cognitive Interference Channel
}

\author{
\IEEEauthorblockN{Stefano Rini$^{\diamond}$, Ernest~Kurniawan$^{\dag}$ and  Andrea Goldsmith$^{\dag}$}
\IEEEauthorblockA{$^{\diamond}$ Technische Universit\"{a}t M\"{u}nchen, Munich, Germany,\\
  $^{\dag}$Stanford University, Stanford, CA, USA\\
Email: \texttt{stefano.rini@tum.de, \{ernestkr,andrea\}@wsl.stanford.edu}}
}

\maketitle

\begin{abstract}
The cognitive interference channel models
cognitive overlay radio systems,
where cognitive radios
overhear the transmission of neighboring nodes.
Capacity for this channel is not known in general.
For the Gaussian case capacity is known in three regimes, usually denoted as the ``weak interference'',  ''very strong interference'' and
``primary decodes cognitive''.
This paper provides a new capacity result, based on rate-splitting of the primary user's message into a public and private part and that generalizes
the capacity results in the ''very strong interference'' and ``primary decodes cognitive'' regimes.
%
%
This result indicates that capacity of the cognitive interference channel not only depends on channel conditions but also the level of cooperation with the primary user.

\end{abstract}

{\IEEEkeywords
cognitive interference channel, capacity , superposition coding, binning, rate-splitting, strong interference.
}

\section{Introduction}
A new generation of smart wireless devices is emerging that can sense and adapt to the surrounding radio environment and this technological development promises to drastically improve the efficiency in using the radio frequency spectrum.
A model that captures the role of cooperation in overlay cognitive radio networks is the \emph{cognitive interference channel} \cite{devroye2005cognitive}.
This channel is obtained from the classic interference channel by providing one of the transmitter, the \emph{cognitive} transmitter, with the message of the
other transmitter,  the \emph{primary} transmitter.
The extra information at the cognitive transmitter models the ability of this node to acquire information about the primary user by exploiting the broadcast
nature of the wireless medium.

The capacity of a cognitive interference channel for both the discrete memoryless case and the Gaussian case remains unknown in general.
However, general outer bounds \cite{maric2005capacity} as well as inner bounds \cite{RTDjournal1} for this channel have been derived.
Capacity is known for the discrete memoryless case in the ``better cognitive decoding'' regime, in which  capacity is achieved
using rate-splitting and superposition coding \cite{RTDjournal1}.
A larger set of capacity results is available for the Gaussian case: here capacity is known in three different regimes.
In the ``weak interference'' regime \cite{WuDegradedMessageSet} capacity is achieved by having the encoders cooperate in transmitting the primary message and by having the primary receiver treat the interference as noise while the cognitive transmitter pre-codes its message against the known interference.
Capacity is known for channels in the ``very strong interference''\cite{maric2005capacity} regime and is achieved by superimposing the cognitive message over the primary message and having both decoders decode both messages.
The last regime in which capacity is known for the Gaussian case is
%
 the ``primary decodes cognitive'' regime \cite{Rini:Allerton2010}.
Here capacity is achieved by pre-coding the cognitive codeword against the interference created by the primary transmission and having the primary receiver decode both the primary and the cognitive codeword.
The primary decoder gains insight over its own message by  decoding the cognitive codeword, since the interference against which the cognitive codeword is pre-coded is indeed the primary codeword.
Capacity for the Gaussian case is also known to within 1 bit/s/Hz and to within a factor of two \cite{Rini:ICC2010}, that is,
a bounded difference  between inner and outer bound
has been established as well as a bounded ratio.

In the following we derive a new capacity result that generalizes the capacity results available for the ``very strong interference'' and the ``primary decodes cognitive'' regimes.
This result is obtained by considering an achievable scheme that includes the capacity achieving schemes in the regimes above as a special cases.
In this scheme the primary message is rate-split into a public and a private part and the private part is then superposed over the public one.
The cognitive message is also superposed over the public primary message and binned against the private primary message.
By determining the optimal rate-splitting between public and private primary message, we obtain capacity in a region that contains both
the ``very strong interference'' and the ``primary decodes cognitive'' regimes.
This result shows, in particular, that the optimal transmission strategy depends not only on the channel condition but also on the
level of cooperation between the cognitive and the primary users.
This is indeed a very interesting results since, in all the previously known capacity results for the Gaussian case, a single transmission scheme achieves capacity in the whole capacity region while,  in this new result, capacity is achieved using two distinct transmission strategies.

%
{\bf Paper Organization:}

The paper is organized as follows:
In Sec. \ref{sec:Channel Model}  we introduce the channel model, the Gaussian cognitive interference channel. In Sec. \ref{sec:Known Results for the G-CIFC} we present some known results for this channel model while in Sec. \ref{sec:Inner Bound} we introduce the inner bound that we will use to prove capacity.
In Sec.  \ref{sec:New Capacity Results} we prove the new capacity result by showing the achievability of an outer bound presented in Sec. \ref{sec:Known Results for the G-CIFC} with the inner bound in Sec. \ref{sec:Inner Bound}.
In \ref{sec:Numerical Simulations} we show the new region in which capacity is derived using numerical simulations.
Sec. \ref{sec:Conclusion} concludes the paper.

\medskip
{\bf  Notation}

$\bullet$
  Let $\Ccal(x)=\log(1+x) $ ,  $\alb=1 - \al$ for $\al \in [0,1]$

$\bullet$
  $\{i...j\}$ indicates the subset of $\Nbb$ between $i$ and $j$,

$\bullet$
  $[i,j]$ indicates the subset of $\Rbb$ between $i$ and $j$.

\section{Channel Model}
\label{sec:Channel Model}

A two-user  Gaussian Cognitive InterFerence Channel (G-CIFC) in standard form \cite[App. A]{RTDjournal2}
is obtained from the classic two-user Gaussian InterFerence Channel  (G-IFC) where the first user is additionally
provided with the message of the second user.

The input/output relationship for this channel is
\eas{
Y_1 & = X_1+   a X_2 + N_1\\
Y_2 & = |b| X_1+  \ X_2 + N_2,
}{\label{eq:channel in/out}}
where $N_i \sim \Ncal(0,1), \ i \in \{1,2\}$ and for $a,b \in \Cbb$.

\begin{figure}
\centering
\includegraphics[width=230pt]{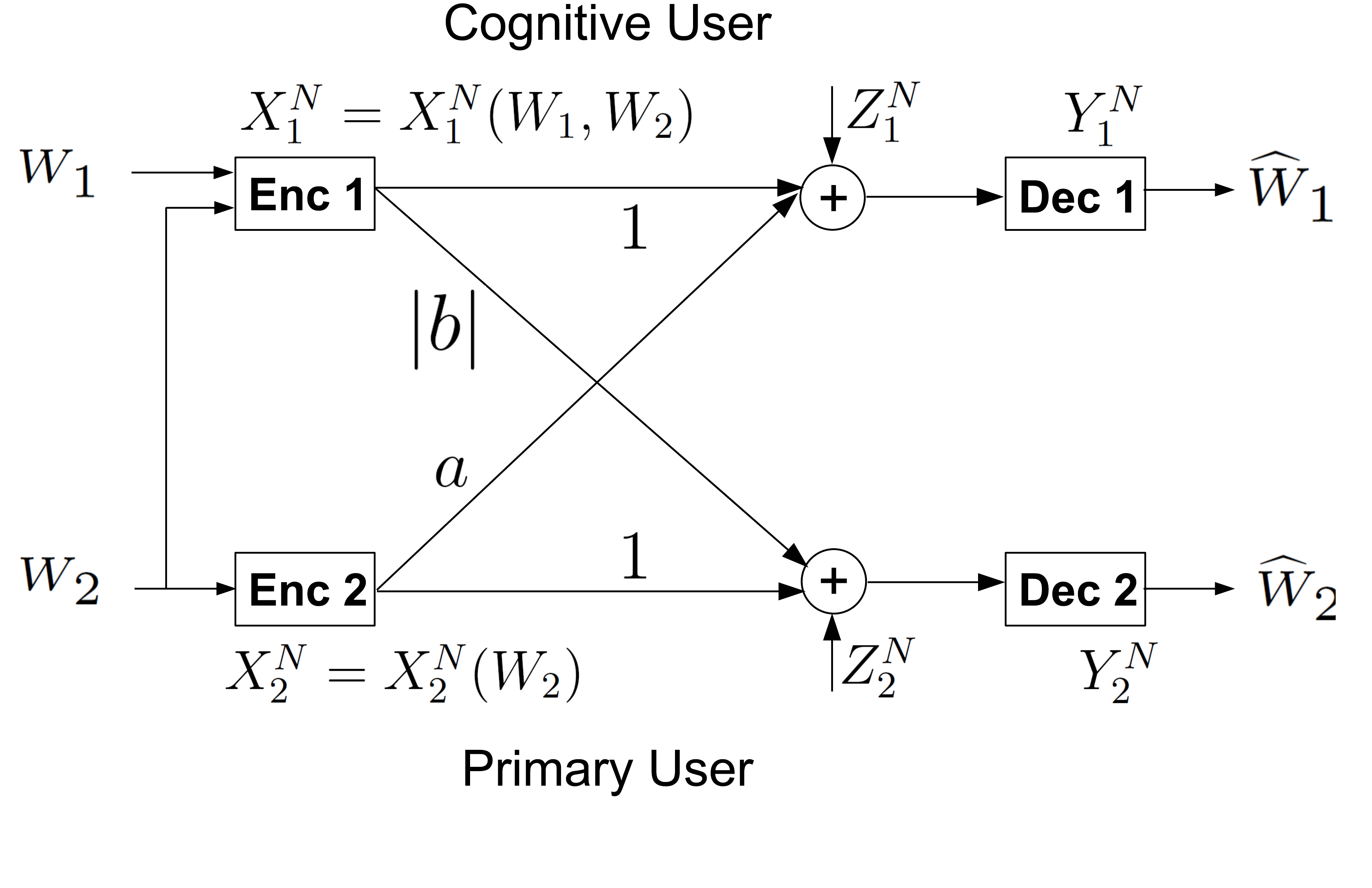}
\vspace{-.8 cm}
\caption{The Gaussian Cognitive InterFerence Channel (CIFC).}
\label{fig:GCIFC}
\vspace{-.5 cm}
\end{figure}

The inputs are additionally subject to a second moment constraint of the form
\eas{
\Ebb[ |X_i|^2] \leq P_i \ \ \ i \in \{1,2\}
}

Encoder $i$ wishes to communicate a message $W_i$ uniformly distributed in $\{1...2^{N R_i}\}$ to decoder $i$ in $N$ channel uses.
The two messages are independent.
Encoder~1, the \emph{cognitive} encoder, is provided with both the message $W_1$ and the message $W_2$.

A rate pair $(R_1,R_2)$ is achievable if there exists two encoding functions
\eas{
X_1 & =f_{X_1} (W_1,W_2) \\
X_2 & =f_{X_2} (W_2),
}
and decoding functions
\eas{
\Wh_1 & =f_{\Wh_1} (Y_1)\\
\Wh_2 & =f_{\Wh_2} (Y_2),
}
such that
\ea{
\Pr \lsb \Wh_1 \neq W_1 \cap \Wh_2 \neq W_2  \rsb  \leq \ep \
}
for any $\ep>0$.

The rate of each user, probability of error and capacity are defined as usual \cite{ThomasCoverBook}.

\section{Known Results for the G-CIFC}
\label{sec:Known Results for the G-CIFC}

We begin by reviewing some known results for the G-CIFC that are relevant for the remainder of the paper.

\begin{thm}{\bf ``Weak Interference'' capacity \cite[Lem. 3.6]{WuDegradedMessageSet}}
\label{th:Weak Interference capacity }
If $|b| < 1$, the capacity of the G-CIFC is the union over $\al \in [0,1]$ of the region
\eas{
R_1 & \leq I(Y_1; X_1 | X_2) \nonumber   \\
    & =  \Ccal(\al P_1) \\
R_2 & \leq I(Y_2; X_2) \nonumber \\
& = \Ccal ( |b|^2 P_1+P_2 + 2 \sqrt{\alb |b|^2 P_1 P_2 })-\Ccal(|b|^2 \al P_1).
}{\label{eq:weak interference capacity}}
\end{thm}

Capacity  in the ``weak interference'' regime is achieved by pre-coding the cognitive codeword against the interference experienced at the cognitive decoder and
treating the interference as noise at the primary decoder.

\begin{thm}{\bf ``Strong interference'' outer bound \cite[Th. 4]{MaricGoldsmithKramerShamai07}}
\label{th:strong int OB}
If
\ea{
|b| \geq 1,
\label{eq:b geq 1}
}the capacity of the G-CIFC is contained in the  union over $\al \in [0,1]$ of the region
\eas{
R_1 & \leq I(Y_1 ; X_1  | X_2) \nonumber \\
    & =  \Ccal(\al P_1)
       \label{eq:strong int OB R1 bound}\\
R_1 + R_2 & \leq  I(Y_2 ; X_1 , X_2)  \nonumber \\
    & = \Ccal ( |b|^2 P_1+P_2 + 2 \sqrt{\alb |b|^2 P_1 P_2 }).
    \label{eq:strong int OB sum rate bound}
}{\label{eq:strong int OB}}
\end{thm}

The rate bound \eqref{eq:strong int OB R1 bound} is a general bound that holds for any G-CIFC and suggests that the largest rate $R_1$
can be achieved by either pre-canceling or decoding $X_2$ at the cognitive receiver.
The sum rate bound \eqref{eq:strong int OB sum rate bound} holds only under condition \eqref{eq:b geq 1}.
The primary receiver, after having decoded its message, can reconstruct the channel output at the cognitive receiver.
This consideration provides an intuitive interpretation of the sum rate bound in \eqref{eq:strong int OB sum rate bound} which suggest
that the primary receiver can decode both messages without loss of optimality.

\begin{thm}{\bf ``Very strong interference'' capacity \cite[Th. 4]{MaricGoldsmithKramerShamai07}}
\label{th:VSI capacity condition}
If condition \eqref{eq:b geq 1} and
\eas{
& (1-|b|^2) P_1 + (|a|^2  -1) P_2  \geq 0 \\
& (1-|b|^2) P_1 + (|a|^2  -1) P_2 \geq  2 \lb |b|-\Re\{a^*\}\rb  \sqrt{\alb P_1 P_2},
}{\label{eq:VSI capacity condition}}
hold, the region in \eqref{eq:strong int OB} is the capacity region.
\end{thm}

\begin{IEEEproof}
The proof for complex channel coefficients can be found in \cite[App. B]{RTDjournal2}.
This result can be improved by noticing that one can restrict $\al \in [0,1]$ in the inner bound to match the strong interference outer bound in Th. \ref{th:strong int OB}.
\end{IEEEproof}

In the \hspace{-1pt}``very strong interference'' \hspace{-1pt}regime, capacity is achieved by having both decoders decode both messages and superimposing the cognitive message over the primary
message.

\begin{thm}{\bf ``Primary decodes cognitive '' capacity \cite[Th. 3.1]{Rini:Allerton2010}}
\label{th:PDC capacity}
If condition \eqref{eq:b geq 1} and
\eas{
\hspace{-3pt}& \hspace{-5pt} P_2 |1-a |b| |^2 (1+ P_1)  \geq (|b|^2-1) (1 + P_1 + |a|^2 P_2) \\
\hspace{-3pt}& \hspace{-5pt} P_2 |1-a |b| |^2  \geq (|b|^2 \!-\!1) (1 + P_1 + |a|^2 P_2 \!-\! 2 \Re\{a\} \sqrt{P_1 P_2})
}{\label{eq:PDC capacity}}
hold, the region in \eqref{eq:strong int OB} is the capacity region.
\end{thm}

In Th. \ref{th:PDC capacity} capacity is achieved by pre-coding the cognitive message against the interference and having the primary receiver decode this codeword as well.

\section{Inner Bound}
\label{sec:Inner Bound}

The largest known inner bound for a general CIFC  is obtained in \cite{rini2009state} while a compact expression for this region is provided in \cite[Sec. IV]{RTDjournal2}.
\cite{RTDjournal2} also provides a series of simpler transmission schemes that are special cases of the most general achievable scheme and
can be expressed using a limited set of parameters.
In the following we consider a transmission scheme that generalizes the capacity achieving schemes in the ``very strong interference''  and the ``primary decodes cognitive''
regimes.

\begin{thm}{\bf Achievable scheme (F) in  \cite[Sec. IV.F]{RTDjournal2}}
\label{th:Achievable scheme (F)}
The following region is achievable in a general CIFC
\eas{
R_1 & \leq  I(Y_1 ; U_{1c}| U_{2c})-I(U_{1c}; X_2 | U_{2c})
\label{eq:inner bound R1 rate bound}\\
R_1 & \leq I(Y_2 ; U_{1c}, X_2 |U_{2c})
\label{eq:inner bound additional R1 rate bound}\\
R_1  + R_2 & \leq 
I(Y_2 ; X_1, X_2)
\label{eq:inner bound sum rate bound 1} \\
R_1  + R_2 & \leq I(Y_2 ; X_2 |U_{1c}, U_{2c}) + I(Y_1 ; U_{1c}, U_{2c}) 
\label{eq:inner bound sum rate bound 2}\\
2 R_1 + R_2
& \leq I(Y_2 ; U_{1c}, X_2 | U_{2c})+ I(Y_1 ; U_{1c}, U_{2c}) \nonumber \\
& \quad \quad - I(U_{1c}; X_2 | U_{2c}),
\label{eq:inner bound additional weird rate bound}
}{\label{eq:inner bound}}
for some distribution
\ea{
P_{U_{2c}} P_{X_2 | U_{2c}} P_{U_{1c} | U_{2c}, X_2} P_{X_1 | U_{1c}, U_{2c}, X_2}.
\label{eq:fact. II}
}
\end{thm}

\begin{figure}
\centering
\includegraphics[width=6 cm]{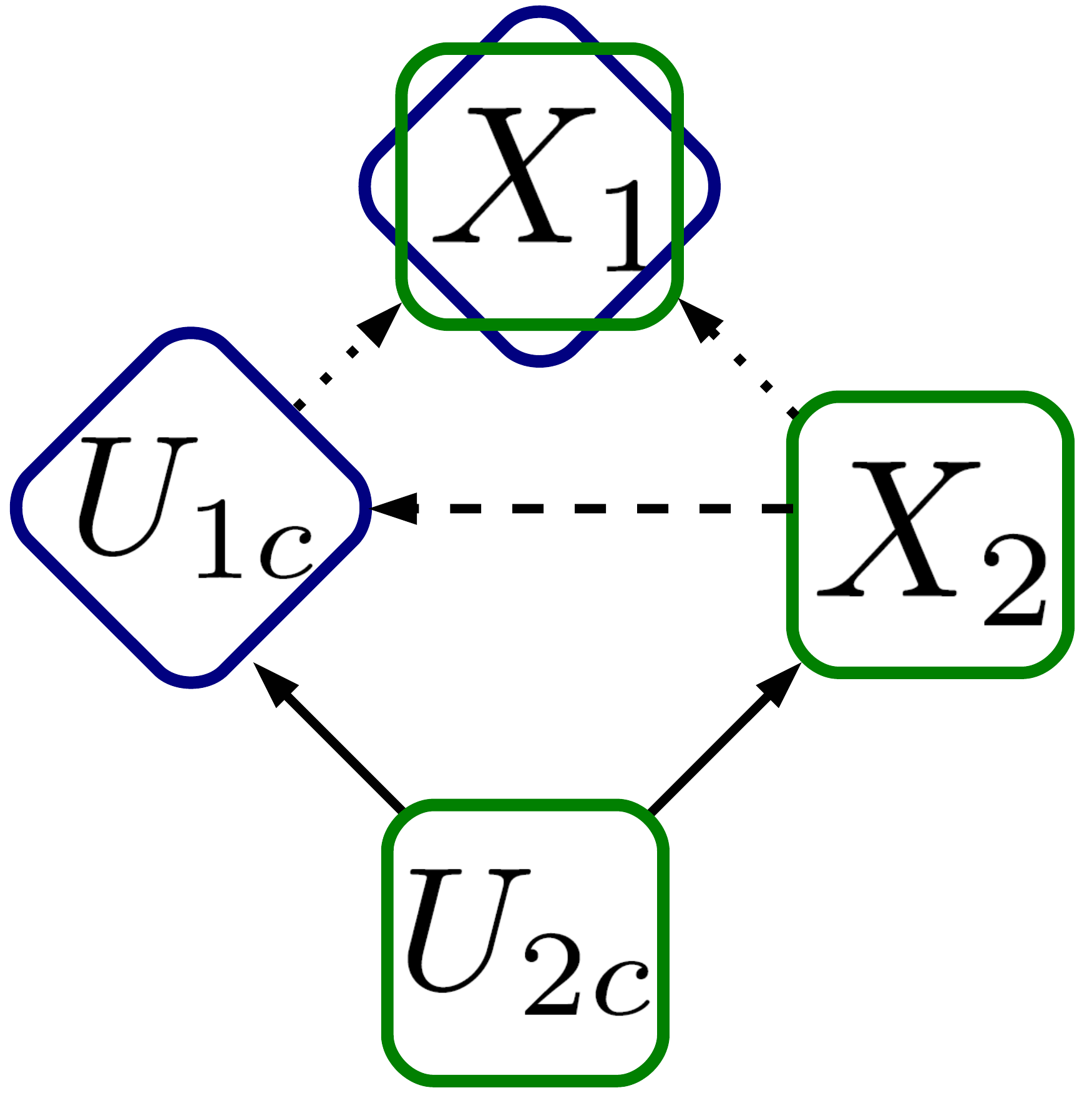}
\vspace{-.2 cm}
\caption{The chain graph representation of the achievable region in Th. \ref{th:Achievable scheme (F)}.}
\label{fig:AchievableRegion}
\vspace{-.5 cm}
\end{figure}

The chain graph \cite{RiniChainGraph} representation of the achievable scheme in Th. \ref{th:Achievable scheme (F)} can be found in Fig. \ref{fig:AchievableRegion}. Each box represents a RV in \eqref{eq:inner bound}, a solid line represents superposition coding, a dashed line binning and a dotted line a deterministic dependence.
Green, square boxes contain part of the message $W_2$ while the blue diamond box the message $W_1$.

For the G-CIFC in \eqref{eq:channel in/out} we consider the following assignment for the Random Variables (RVs) in \eqref{eq:fact. II}:
\eas{
X_{i} & \sim \Ncal_{\Cbb}(0,1) \quad i \in \{1c,2c,2p\}\\
X_2 & = \sqrt{P_2 \be } X_{2c} + \sqrt{P_2 \beb } X_{2p} \\
X_1 & = \sqrt{\al P_1} X_{1c} + \sqrt{\f {\alb P_1} {P_2}} X_2 \\
U_{1c} & =\sqrt{\al P_1} X_{1c} + \la_{\rm Costa \ 1} X_{2p}  \\
\la_{\rm Costa \ 1}& = \f {\al P_1} {\al P_1 + 1} a \sqrt{P_2},
}{\label{eq:assigment gaussian}}

\section{New Capacity Results}
\label{sec:New Capacity Results}

We now show how the outer bound in Th. \ref{th:strong int OB} can be achieved using the inner bound of Th.
\ref{th:Achievable scheme (F)} by optimally choosing the rate-splitting between the public and the private part.
We begin by  considering the case where only superposition coding is employed.

%

\begin{lem}{\bf Partial achievability of the ``strong interference'' outer bound with superposition coding \cite{maric2005capacity}}
\label{th:partial superposition}
The  ``strong interference'' outer point for $\al=\al'$ is achievable if $|b|\geq 1$ and
\ea{
& (1-|b|^2)P_1 + (a^2-1 )P_2 \geq \nonumber \\
& \quad  \quad  \quad \quad  2 \lb |b| -\Re\{a^*\}\rb \sqrt{ \alb' P_1 P_2}.
\label{eq:partial superposition}
}

\end{lem}

\begin{IEEEproof}
When fixing the rate of the private primary message to zero in \eqref{eq:inner bound}, which is equivalent to setting $\be=1$ in \eqref{eq:assigment gaussian},
and for $|b| \geq 1$, the rate bounds \eqref{eq:inner bound additional R1 rate bound} and \eqref{eq:inner bound additional weird rate bound} can be dropped.
With this choice, the outer bound point in Th. \ref{th:strong int OB} for  $\al=\al'\in [0,1]$ is achievable when
\eas{
 I(Y_1 ; U_{1c},  U_{2c}) & \geq I(Y_2 ; X_1 X_2) \quad \IFF \\
 I(Y_1 ; X_1,  X_2) & \geq I(Y_2 ; X_1 X_2)
}
which corresponds to the condition in \eqref{eq:partial superposition} for the assignment in \eqref{eq:assigment gaussian} with $\be=1$.
\end{IEEEproof}
The capacity result in Th. \ref{th:VSI capacity condition} is obtained by imposing condition \eqref{eq:partial superposition} for all $\al' \in [0,1]$.

We now consider the case when only binning is employed in the achievable scheme of Th. \ref{th:Achievable scheme (F)}.

\begin{thm}{\bf Partial achievability of the ``strong interference'' outer bound with binning \cite{Rini:Allerton2010}}
\label{th:partial binnnig}
The  ``strong interference'' outer point for $\al=\al'$ is achievable if $|b|\geq 1$ and
\ea{
& P_2 \lb 1-a|b| \rb^2 \lb  \al P_1+1 \rb
\label{eq:partial binnnig} \\
& \quad \quad - (|b|^2-1)(P_1 + |a|^2 P_2 + 2 a  \sqrt{\alb' P_1 P_2} +1 ) \geq 0. \nonumber
}
\end{thm}

\begin{IEEEproof}
When fixing the rate of the common primary  message to zero in \eqref{eq:inner bound}, which is equivalent to setting $\be=0$ in \eqref{eq:assigment gaussian},
and  for $|b| \geq 1$, the rate bounds \eqref{eq:inner bound additional R1 rate bound} and \eqref{eq:inner bound additional weird rate bound} can be dropped.
With this choice, the outer bound point in Th. \ref{th:strong int OB} for  $\al=\al'\in [0,1]$ is achievable when
\eas{
I(Y_2 ;  X_2 | U_{1c}) + I(Y_1 ; U_{1c}) & \geq  I(Y_2; X_1 , X_2) \quad \IFF \\
I(Y_1; U_{1c}) & \geq I(Y_2; U_{1c})
}
which corresponds to the condition in \eqref{eq:partial binnnig} for the assignment in \eqref{eq:assigment gaussian} with $\be=0$.
\end{IEEEproof}

With the aid of  Lem. \ref{th:partial superposition} and Lem. \ref{th:partial binnnig}, we now show the achievability of the ``strong interference''
outer bound for $|b| \geq 1$ using the inner bound in Th. \ref{th:Achievable scheme (F)}.

\begin{thm}{\bf A new capacity result}
\label{th:A New Capacity Result}

Let $\lnone (i)\rabs_{\al'= \gamma}$ indicates that condition $(i)$ holds  for the assignment $\al'= \gamma$ and
define
\ea{
\widetilde{\al}=\max \lcb 0 , \min \lcb 1, \f{ (|a|^2 -1) P_2+(1-|b|^2) P_1}{2 (\Re\{a^*\}-|b|) \sqrt{P_1 P_2}} \rcb \rcb  .
\label{eq:condition 1 al when superposition is zero}
}
If
\ea{
\lnone \eqref{eq:partial binnnig}\rabs_{\al'=0} , \quad \lnone \eqref{eq:partial superposition}\rabs_{\al'=1} , \quad \lnone \eqref{eq:partial binnnig}\rabs_{\al'=\widetilde{\al}}
\label{th:A New Capacity Result condition 1}
}
or
\ea{
\lnone \eqref{eq:partial binnnig}\rabs_{\al'=1} , \quad \lnone \eqref{eq:partial superposition}\rabs_{\al'=0}, \quad \lnone \eqref{eq:partial binnnig}\rabs_{\al'=\widetilde{\al}}
\label{th:A New Capacity Result condition 2}
}
the region in \eqref{eq:strong int OB} is the capacity region.
\end{thm}

\begin{IEEEproof}
We now seek to extend the results of Lem. \ref{th:partial superposition} and Lem. \ref{th:partial binnnig} to all the $\al \in [0,1]$.
Since we can show achievability under condition \eqref{eq:partial superposition} and \eqref{eq:partial binnnig}, we only need to focus on the range
of $\al$ for which neither of these conditions hold.
In particular, \eqref{eq:partial superposition} is linear in $\al$, so if it holds for $\al_1$ and $\al_2$, then it holds for the whole interval $[\al_1,\al_2]$.
Similarly, \eqref{eq:partial binnnig} is quadratic and concave in $\sqrt{\alb}$, so if it holds for  $\al_1$ and $\al_2$, then it holds for the whole interval $[\al_1,\al_2]$.

To match the inner bound in Th. \ref{th:Achievable scheme (F)} with the assignment in \eqref{eq:assigment gaussian}
and the outer bound in Th.   \ref{th:strong int OB} for $\be\geq 1$ we need equations \eqref{eq:inner bound additional R1 rate bound}, \eqref{eq:inner bound sum rate bound 2} and \eqref{eq:inner bound additional weird rate bound} to be redundant, that is
\eas{
I(Y_2; U_{1c}, X_2 | U_{2c}) & \geq I(Y_1 ; U_{1c} | X_2, U_{2c} )
\label{eq:conditions new capacity 1}\\
I(Y_1; U_{1c}, U_{2c}) & \geq I(Y_2; U_{1c}, U_{2c})
\label{eq:conditions new capacity 2}\\
I(Y_1; U_{2c}) & \geq  I(Y_2; U_{2c}).
\label{eq:conditions new capacity 3}
}{\label{eq:conditions new capacity}}

Condition \eqref{eq:conditions new capacity 3} can be rewritten as
\ea{
& \f {|a|^2P_2 + P_1 + 2 \Re\{a^*\} \sqrt{\alb P_1 P_2}+1}{ \al P_1 + \beb \labs \sqrt{\alb P_1} + a \sqrt{ P_2} \rabs^2 +1}
\label{eq:gaussian condition sum rate redundant}  \\
& \quad \quad  \geq \f{|b|^2 P_1 + P_2 + 2 |b|\sqrt{\alb P_1 P_2}+1 }{ |b|^2\al P_1 + \beb  \labs \sqrt{ |b|^2\alb P_1} + \sqrt{P_2} \rabs^2 +1}.
\nonumber
}
Note that condition \eqref{eq:gaussian condition sum rate redundant} holds only for $\be=0$ in the ``strong interference''  but outside the ``very strong interference'' regime : this means that when condition \eqref{eq:partial superposition}
does not hold, one can hope to achieve the outer bound only with the choice $\be=0$.
With this observation we conclude that we can achieve capacity using $\be=0$ for a subset of the $\al$ while using $\be=1$ for the remaining subset,
that is either condition \eqref{eq:partial superposition} or \eqref{eq:partial binnnig} must hold for any $\al \in [0,1]$.
Since both conditions have at most two zeros in $\al'$, the above condition is satisfied when either
\begin{itemize}
  \item one condition holds in both zero and one, or
  \item one condition holds in zero and is in $\al'=\widetilde{\al}$, the other holds in one and in $\al'=\widetilde{\al}$.
\end{itemize}
For simplicity, one can chose $\widetilde{\al}$ to be the $\al'$, for which condition \eqref{eq:partial superposition}
holds with equality as in \eqref{eq:condition 1 al when superposition is zero}.
\end{IEEEproof}
%
%
In the above proof, the optimal transmission strategy is obtained by either a private public message on a private one, depending on the cooperation level between the transmitters.
%
%
This is somewhat surprising as one would expect some rate advantage from primary message private and a part public.
The key intuition here is provided by \eqref{eq:gaussian condition sum rate redundant}: outside the ``very strong interference'' regime there is a
rate penalty in decoding the primary message at the cognitive decoder at some rates.
When such penalty exists, the best thing to do is to set the rate of the private cognitive message to zero.
%
%
Note that this may not be the case when considering an assignment different from \ref{eq:assigment gaussian}: in \cite{Rini:Allerton2010} it is shown that
partial interference cancelation, i.e. $\la \neq \la_{\rm Costa \ 1}$ in \eqref{eq:gaussian condition sum rate redundant}, can yield large achievable regions then full interference cancelation.
%

%
%

\begin{figure}
\centering
\includegraphics[width=252pt]{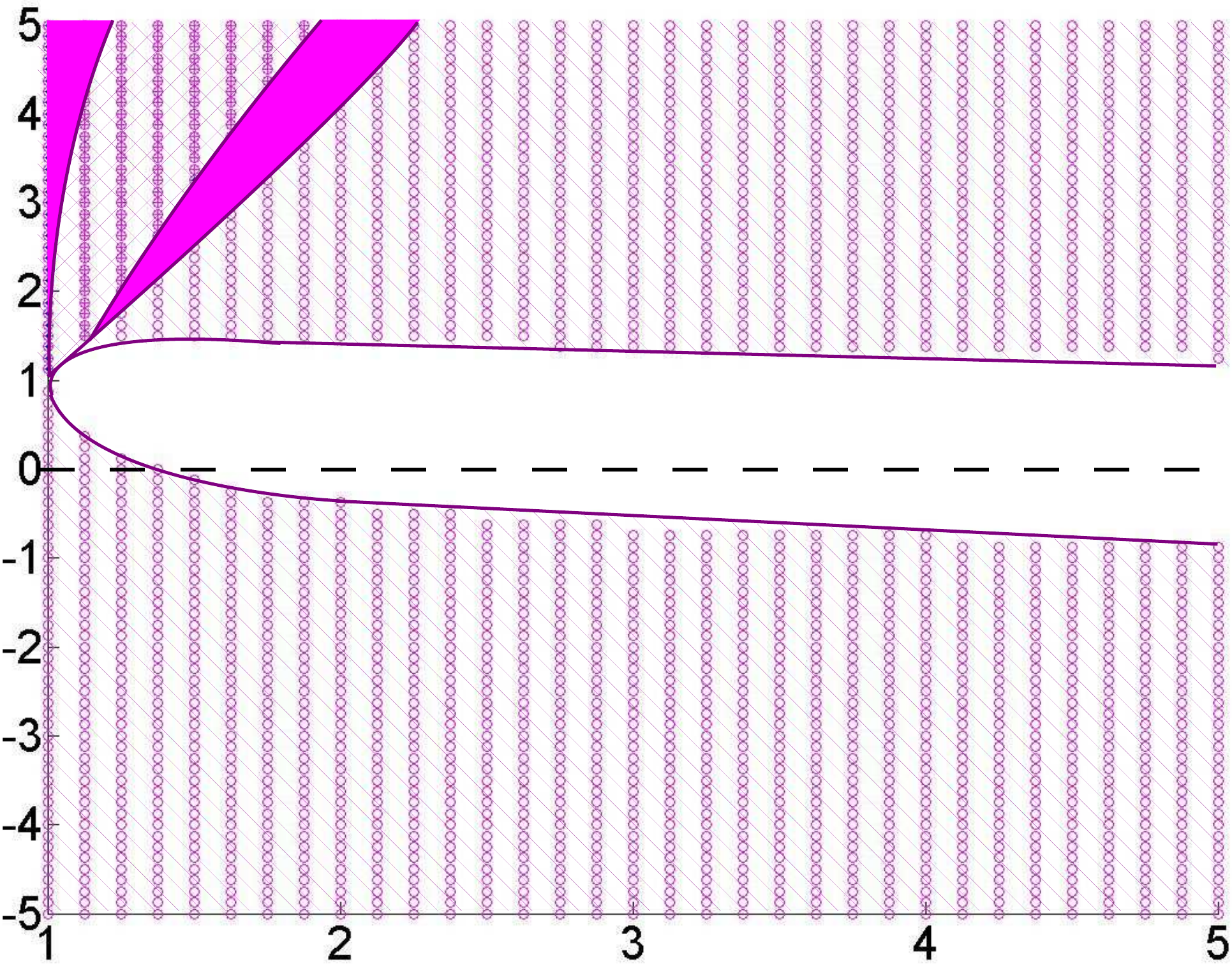}
\vspace{-.4 cm}
\caption{The region where \eqref{eq:partial binnnig} for $\al'=0$ (single hatched) holds, where \eqref{eq:partial superposition} for $\al'=1$ (cross hatched)
holds and where \eqref{th:A New Capacity Result condition 1} holds (solid color),  for $P_1=10$, $P_2=1$ and $a \times |b| \in [-5,5]\times[1,5]$.}
\label{fig:condition11}
\vspace{-.2 cm}
\end{figure}

\begin{figure}
\centering
\includegraphics[width=252pt]{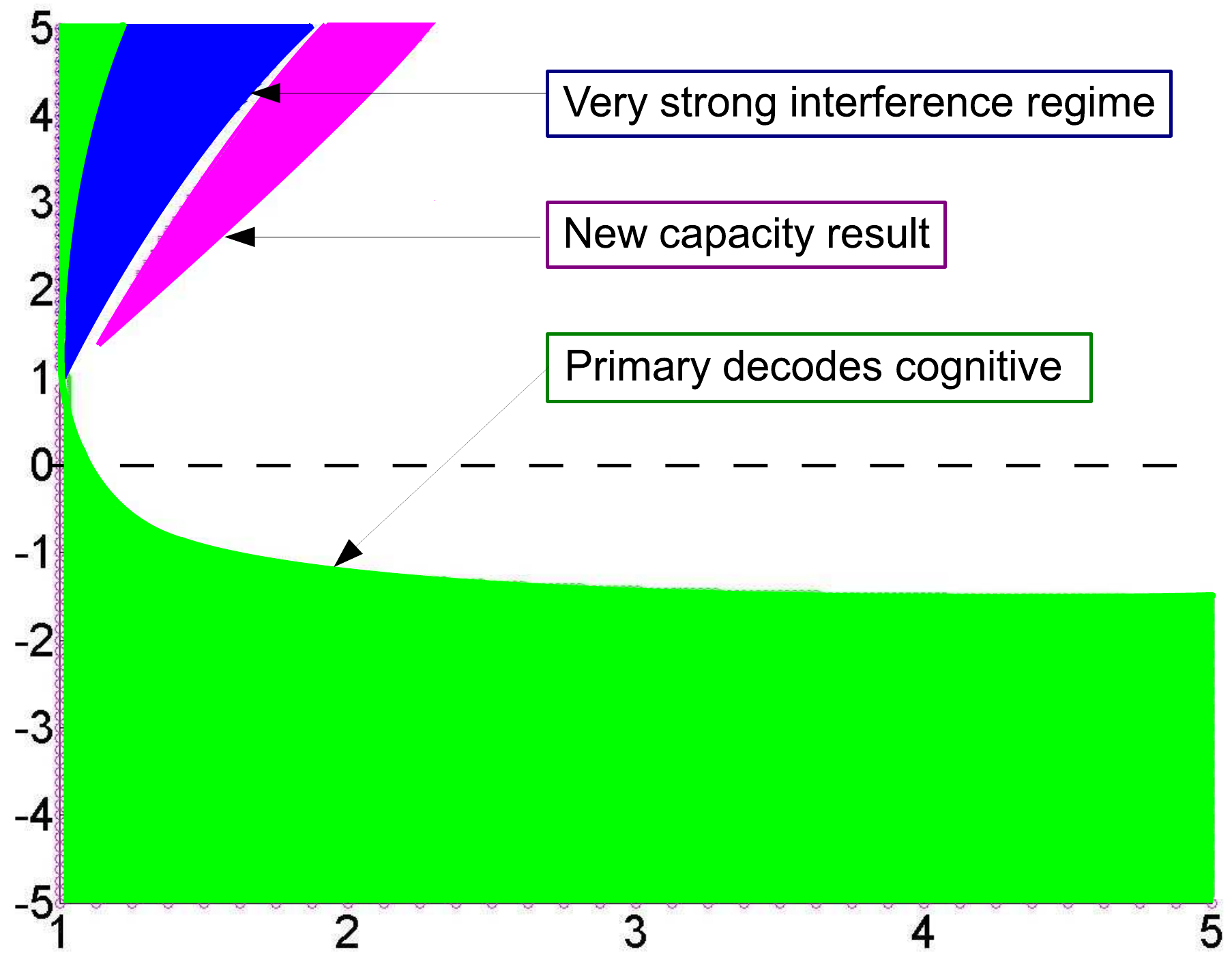}
\vspace{-.8 cm}
\caption{The primary decoder ``very strong interference'' capacity region (blue hatched) holds, where \eqref{eq:partial superposition} for $\al'=1$ (cross hatched)
holds and where \eqref{th:A New Capacity Result condition 1} holds (solid color),  for $P_1=10$, $P_2=1$ and $a \times |b| \in [-5,5]\times[1,5]$.}
\label{fig:condition12}
\vspace{-.2 cm}
\end{figure}

\section{Numerical Results}
\label{sec:Numerical Simulations}

In this section we present some numerical rusults to illustrate the capacity region associated with Th. \ref{th:A New Capacity Result}.

We begin by illustrating the result in \eqref{th:A New Capacity Result condition 1} with Fig. \ref{fig:condition11}: here we plot the region where \eqref{eq:partial binnnig} holds for $\al'=0$, \eqref{eq:partial superposition} holds for $\al'=1$ and finally where \eqref{th:A New Capacity Result condition 1} holds.
Both conditions \eqref{eq:partial binnnig} for $\al'=0$ and  \eqref{eq:partial superposition} for $\al'=1$ are sufficient conditions for
\eqref{th:A New Capacity Result condition 1} to holds.

In Fig. \ref{fig:condition12} we present the improvement on the known capacity region that is provided
by condition \eqref{eq:conditions new capacity 1}. In this figure we represent
the ``very strong interference'' regime  of Th. \ref{th:VSI capacity condition} and the ``primary decodes cognitive'' regime of Th. \ref{th:PDC capacity} together
with the new capacity result in Th. \ref{th:A New Capacity Result}.

\begin{figure}
\centering
\includegraphics[width=252pt]{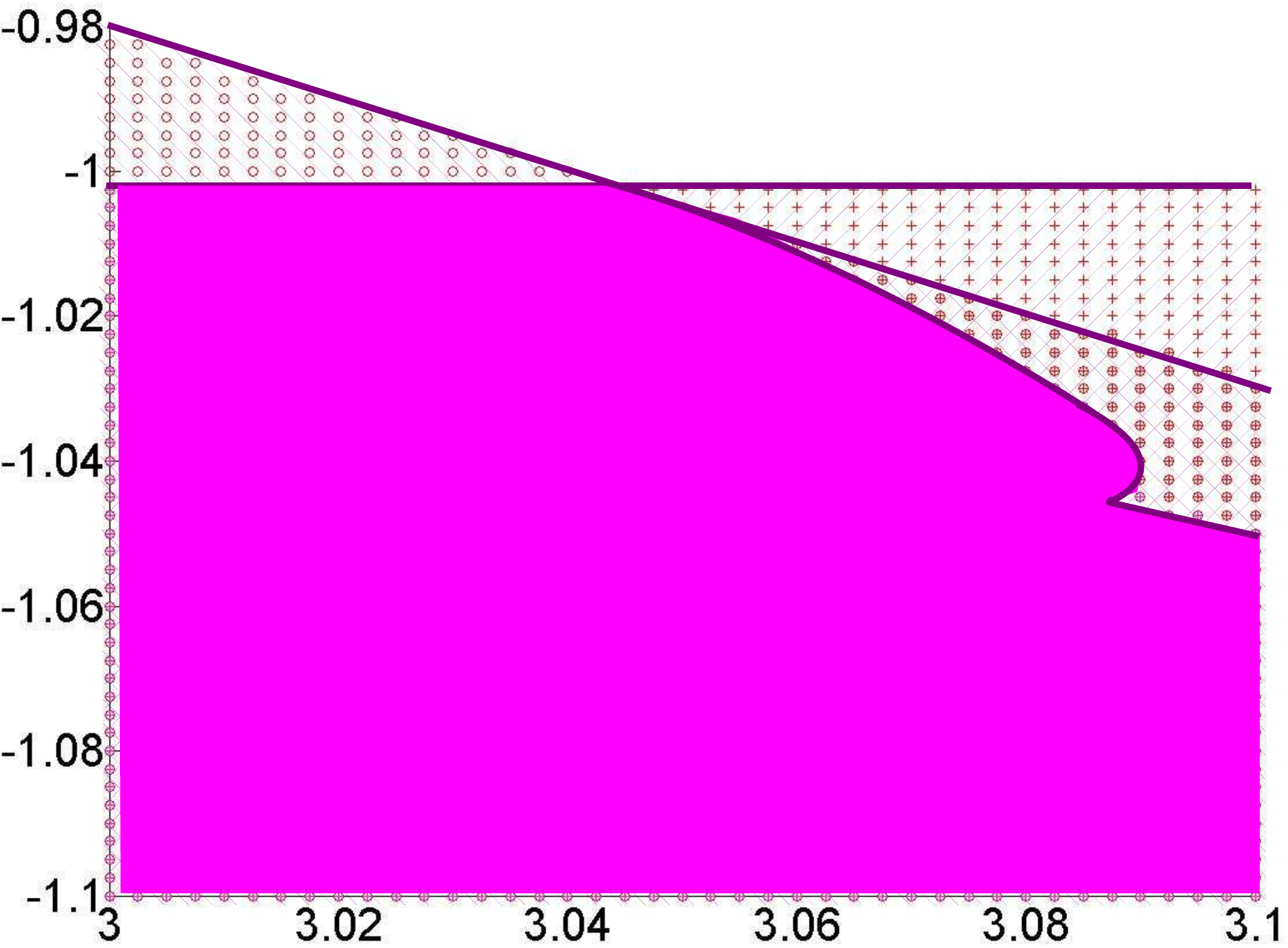}
\vspace{-.8 cm}
\caption{The region where \eqref{eq:partial binnnig} for $\al'=1$ (left hatched) holds, where \eqref{eq:partial superposition} for $\al'=0$ (right hatched)
holds and where \eqref{th:A New Capacity Result condition 1} holds (solid color),  for $P_1=10^{-3}$, $P_2=1$ and $a \times |b| \in [-1.1,-1]\times[3,3.1]$.}
\label{fig:condition21}
\vspace{-.2 cm}
\end{figure}

\begin{figure}
\centering
\includegraphics[width=252 pt]{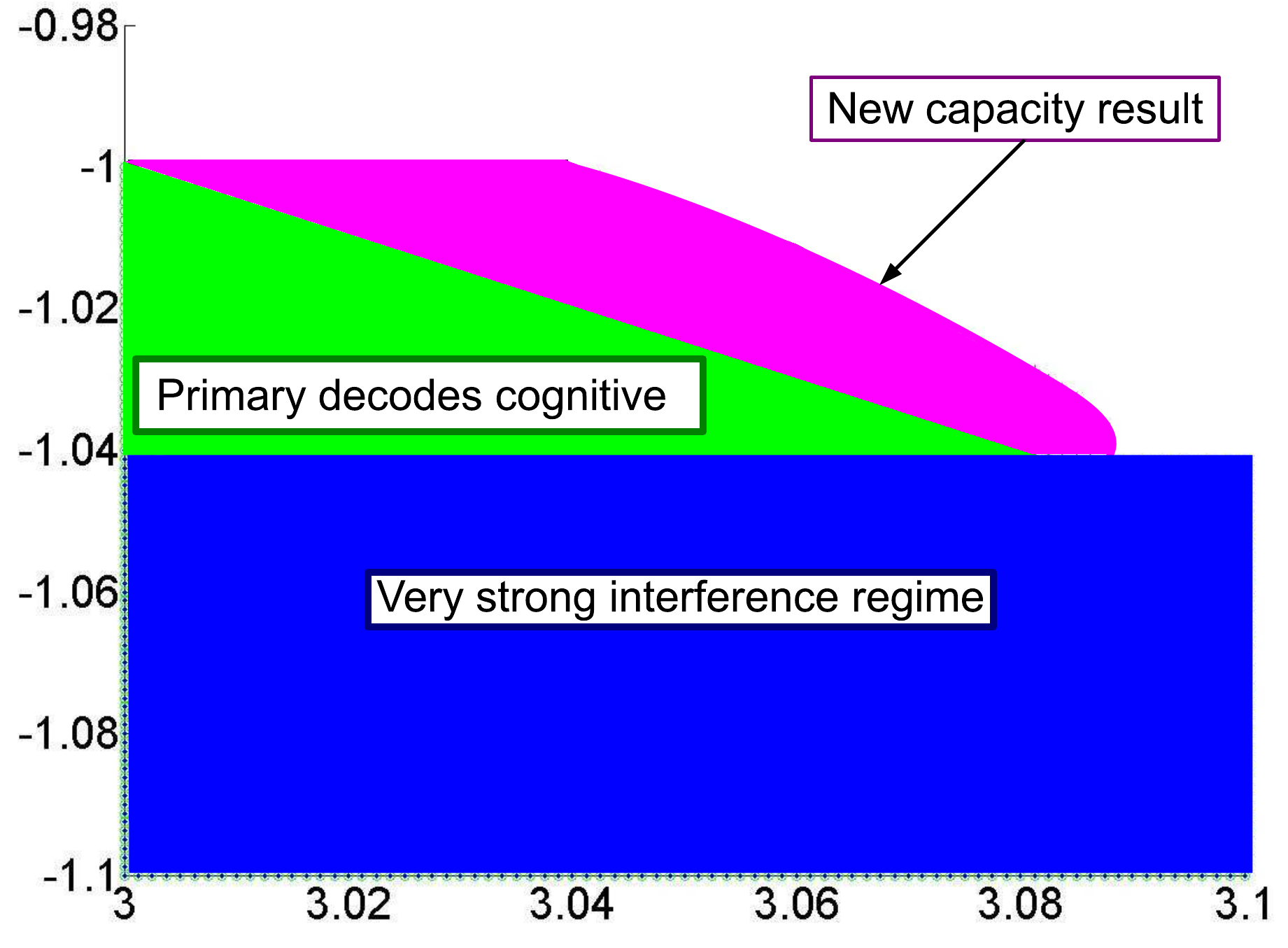}
\vspace{-.8 cm}
\caption{The primary decoder ``very strong interference'' capacity region (blue hatched) holds, where \eqref{eq:partial superposition} for $\al'=1$ (cross hatched)
holds and where \eqref{th:A New Capacity Result condition 1} holds (solid color),  for $P_1=10^{-3}$, $P_2=1$ and $a \times |b| \in [-1.1,-1]\times[3,3.1]$.}
\label{fig:condition22}
\vspace{-.2 cm}
\end{figure}

Fig. \ref{fig:condition21} and Fig. \ref{fig:condition22} are the analogous of Fig. \ref{fig:condition11}  and Fig. \ref{fig:condition12} for condition \eqref{th:A New Capacity Result condition 2}.

\section{Conclusion}
\label{sec:Conclusion}

In this paper we derive a new capacity result for the cognitive interference channel, a classic interference channel where the first transmitter
is additionally provided with the message of the second user.
This new capacity result is obtained by generalizing the capacity proof for the ``very strong interference'' regime, where superposition coding achieves capacity,
and for the ``primary decodes cognitive'' regime, where binning is optimal.
Although this result improves on the class of channels for which capacity is known, the complete characterization of the capacity of this channel is still an open
problem.

\section*{Acknowledgment}
The authors would like to thank Professor Gerhard Kramer for valuable discussions
and precious insights on the problem.

\bibliographystyle{IEEEtran}
\bibliography{steBib1}

\end{document}